\documentclass[preprint, journal]{vgtc}                

\usepackage{mathptmx}
\usepackage{graphicx}
\usepackage{times}
\usepackage{subfigure} 

\usepackage{flushend}

\usepackage[bookmarks,linkcolor=black]{hyperref} 
\hypersetup{
  pdfauthor = {},
  pdftitle = {},
  pdfsubject = {},
  pdfkeywords = {},
  colorlinks=true,
  linkcolor= black,
  citecolor= black,
  pageanchor=true,
  urlcolor = black,
  plainpages = false,
  linktocpage
}

\onlineid{0}

\vgtccategory{Research}

\vgtcinsertpkg

\title{Gatherplots: Generalized Scatterplots for Nominal Data }

\author{Deokgun Park, Sung-Hee Kim, and Niklas Elmqvist,
  \textit{Senior Member, IEEE}}

\authorfooter{
\item Deokgun Park, Sung-Hee Kim, and Niklas Elmqvist are with Purdue
  University. E-mail: \{park573, kim731, elm\}@purdue.edu.
}

\shortauthortitle{Park \MakeLowercase{et al.}: Gatherplots}

\preprinttext{}

\teaser{
  \centering
  \includegraphics[width=6in]{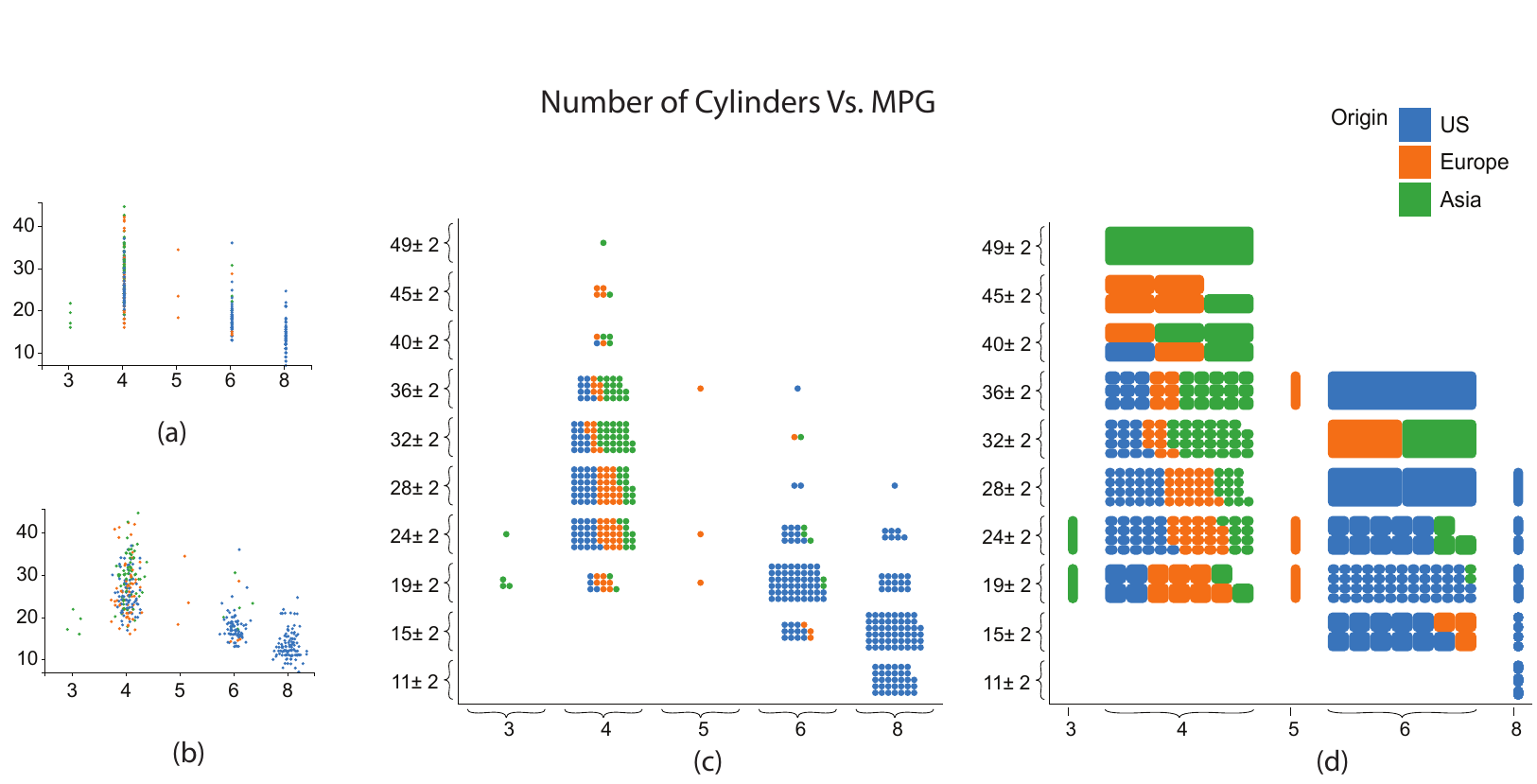}
  \caption{Multiple plots showing a relation between the number of
    cylinders vs.\ gas mileage (MPG) of a car datset.
    Item colors encode the origin of the car.
    The scatterplot in (a) suffers from overplotting on the Y axis
    because the number of cylinders assigned to the X axis is an
    ordinal variable.
    To avoid this, the scatterplot in (b) uses random jittering.
    However, even though the overall sizes of groups with different
    number of cylinders are now visible, estimating the distribution
    of origins is difficult.
    To alleviate this, the gatherplot in (c) partitions the graphical
    axes into intervals and stacks points into groups for each
    interval.
    The brackets on the Cartesian axes are used to indicate these
    intervals in the data.
    In (d), cars having cylinders other than 4 and 6 have been folded,
    and which represents a data slicing operation.
    The visual marks for cars with 4 and 6 cylinders have been made
    variable-sized so that they fill the available space in each
    segment, thereby aiding assessment of ratios for different origins
    within each group.}
  \label{fig:teaser}
}

\abstract{Overplotting of data points is a common problem when
  visualizing large datasets in a scatterplot, particularly when
  mapping nominal dimensions to one of the scatterplot axes.
  Transparency, aggregation, and jittering have previously been
  suggested to address this issue, but these solutions all have
  drawbacks for assessing the data distribution in the plot.
  We propose gatherplots, an extension of scatterplots that eliminates
  overplotting, particularly for nominal variables.
  In gatherplots, every data point that maps to the same position
  coalesces to form a stacked entity, thereby making it easier to
  compare the absolute and relative sizes of data groupings.
  The size and aspect ratio of data points can also be changed
  dynamically to make it easier to compare the composition of
  different groups.
  Furthermore, several embedded interaction techniques support slicing
  and dicing the gatherplot by pivoting on particular dimensions,
  ranges, and values in the dataset.
  Our evaluation shows that gatherplots enable users from the general
  public to judge the relative portion of subgroups more quickly and
  more correctly than when using conventional scatterplots with
  jittering.
  Furthermore, a review conducted by a group of visualization experts
  evaluated and commented on the gatherplot design.}


\begin{document}


\firstsection{Introduction}

\maketitle


Scatterplots---one of the most widely used types of statistical
graphics~\cite{Cleveland1988, Elmqvist2008, Utts1996}---are commonly
used to visualize two continuous variables using visual marks mapped
to a two-dimensional Cartesian space, where the color, size, and shape
of the marks can represent additional dimensions.
However, scatterplots are so-called \textit{overlapping}
visualizations~\cite{Fekete2002} in that the visual marks representing
individual data points may begin to overlap each other in screen space
in situations when the marks are large, when there is insufficient
screen space to fit all the data at the desired resolution, or simply
when several data points share the same value.
The latter is particularly problematic for discrete variables with
small domains, such as for nominal and ordinal
data~\cite{Stevens1946}, due to the increased incidence of shared
values.
This kind of overlap is known as \textit{overplotting} (or
\textit{overdrawing}) in visualization, and is problematic because it
may lead to data points being entirely hidden by other points, which
in turn may lead to the viewer making incorrect assessments of the
data.
In particular, the issue with small-domain discrete variables has led
to scatterplots being almost exclusively used for continuous
variables; Figure~\ref{fig:teaser}(a) shows an example of what happens
if a data dimension mapped to an axis has this discrete property.

\begin{figure}[htb]
 \centering
 \includegraphics[width=3in]{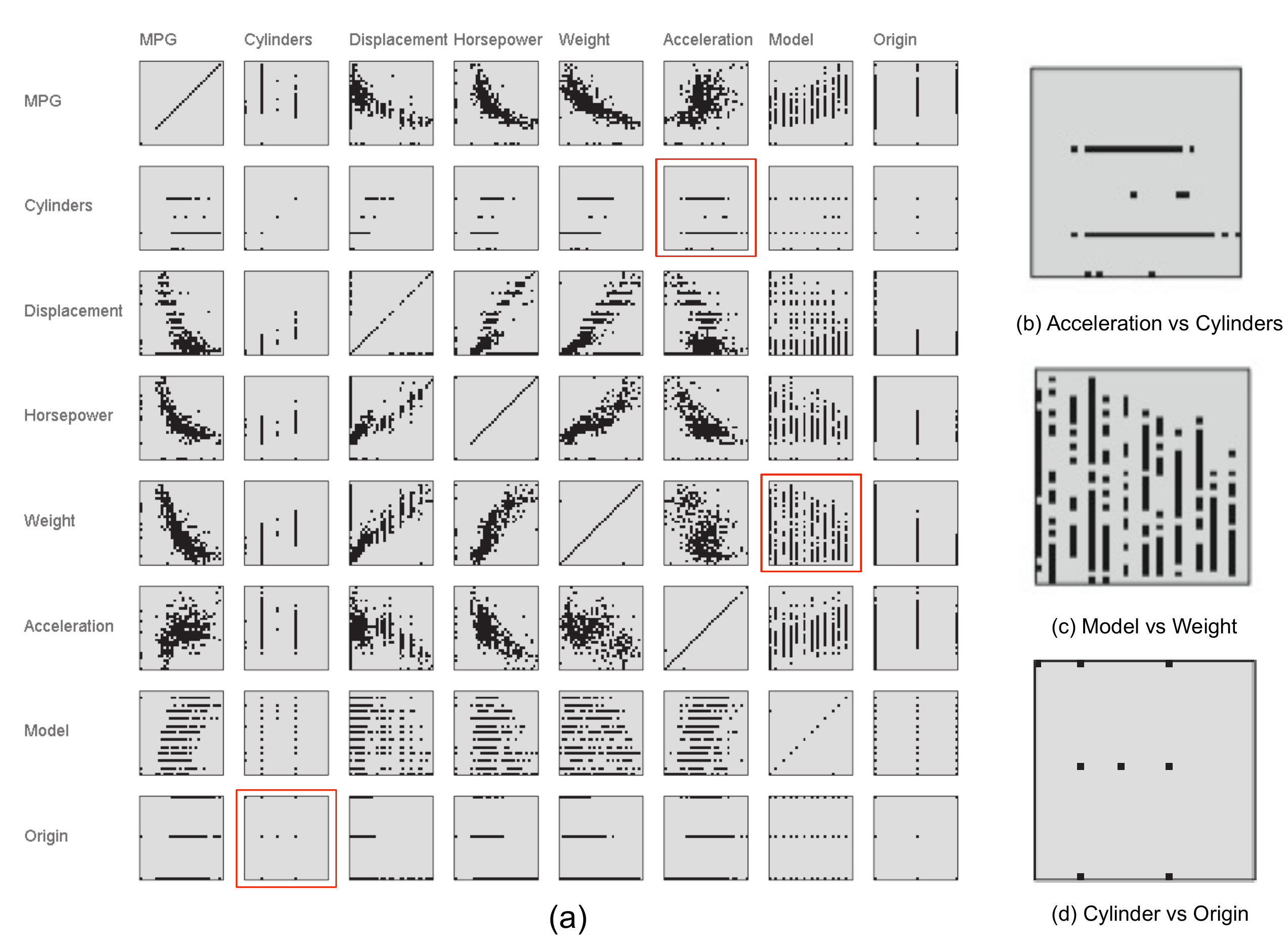}
 \caption{Limitations of scatterplots for managing discrete variables.
   (a) The scatterplot matrix for 7-dimensional car dataset.
   (b) The acceleration vs.\ cylinders scatterplot looks like a
   horizontal line because the ordinal cylinders axis has been mapped
   onto the graphical Y axis.
   (c) The model vs.\ weight scatterplot shows vertical lines due
   to the model year behaving as a discrete ordinal
   value, which causes overplotting.
   (d) The cylinders vs.\ origin scatterplot shows dotted patterns
   because discrete dimensions have been mapped onto both the X and Y
   axis.}
 \label{fig:scatterplot-matrix}
\end{figure}

However, realistic multidimensional datasets often contain a
combination of both continuous and discrete data dimensions, and even
if a dataset is entirely continuous, the physical limitations of
computer screens means that overplotting may (and most likely will)
still occur even if no value overlap exists in the data
(Figure~\ref{fig:scatterplot-matrix}).
Several approaches have been proposed to address this
problem~\cite{Ellis2007}, the most prominent being transparency,
clustering, and jittering.
The first of these, transparency, does not so much address the problem
as sidestep it by making the visual marks semi-transparent so that an
accumulation of overlapping points in the same are still visible.
However, this will not scale well for large datasets, and also causes
blending issues if color is used to encode additional variables.
Clustering, on the other hand, attempts to organize overlapping marks
into visual groups that summarize the distribution~\cite{Fua1999,
  Mayorga2013}, but increases the complexity of the scatterplot.
Finally, jittering perturbs visual marks using a random
displacement~\cite{Trutschl2003} so that no mark falls on the exact
same screen location as any other mark (Figure~\ref{fig:teaser}(b)),
but this approach is still prone to overplotting for large datasets.
Jittering also introduces uncertainty in the data that is not aptly
communicated by the scatterplot since marks will no longer be placed
at their true location on the Cartesian space.

In this paper, we propose the concept of \textit{gathering} as an
alternative to scattering and jittering, and then show how we can use
this visual transformation to define a novel visualization technique
called a \textit{gatherplot}.
Gathering is a generalization of the linear mapping used by
scatterplots, and works by partitioning the graphical axis into
segments based on the data dimension and then organizing points into
\textit{stacked groups} for each segment that avoids overplotting.
This means that the gather operation relaxes the continuous spatial
mapping traditionally used for a graphical axis; instead, each
discrete segment occupies a certain amount of screen space that is all
defined to map to the exact same data value.
This is communicated using graphical brackets on the axis that shows
the value or interval for each segment (Figure~\ref{fig:teaser}(c)).

The gatherplot technique, then, is merely a scatterplot where the
gather transformation is used on one or both of the graphical axes.
Additional data dimensions can be used to cluster together different
points within a stacked group; Figure~\ref{fig:teaser} shows how the
origin of cars, communicated using color, is also used to organize
these marks into discrete groupings.
Furthermore, if the user is trying to assess relative proportions
rather than absolute numbers, the aspect ratio of the visual marks in
each stacked group can be changed independently to fill the available
space (Figure~\ref{fig:teaser}(d)).
Because we define a common model for scatterplots, jitterplots, and
gatherplots alike, our prototype implementation makes it easy to
transition freely between them.

The contributions of our paper are the following: (1) the concept of
the gather visual transformation as a generalization of linear visual
mappings; (2) the gatherplot technique, an application of the gather
operation to scatterplots to solve the overplotting problem; and (3)
results from both a crowdsourced graphical perception evaluation
studying effectiveness of gatherplots compared to jitterplots as well
as an expert review involving multiple visualization experts using the
technique for in-depth data analysis.
In the remainder of this paper, we first review the literature on
statistical graphics and overplotting.
We then present the gather operation and use it to define gatherplots.
This is followed by our crowdsourced evaluation and expert review.
We close with implementation notes, conclusions, and our future plans.

\section{Background}

Our goal with gatherplots is to generalize scatterplots to a
representation that maintains its simplicity and familiarity while
eliminating overplotting.
With this in mind, below we review prior art that generalizes
scatterplots for mitigating overplotting.
We also discuss related visualization techniques specifically designed
for nominal variables.

\subsection{Characterizing Overplotting}

While there are many ways to categorize visualization techniques, a
particularly useful classification for our purposes is one introduced
by Fekete and Plaisant~\cite{Fekete2002}, which splits visualization
into two types:

\begin{itemize}

\item\textbf{Overlapping visualizations:} These techniques enforce no
  layout restrictions on visual marks, which may lead to them
  overlapping on the display and causing occlusion.
  Examples include scatterplots, node-link diagrams, and parallel
  coordinates.

\item\textbf{Space-filling visualizations:} A visualization that
  restricts layout to fill the available space and to avoid
  overlap.
  Examples include treemaps, adjacency matrices, and choropleth maps.

\end{itemize}

Fekete and Plaisant~\cite{Fekete2002} investigated the overplotting
phenomenon for a 2D scatterplot, and found that it has a significant
impact as datasets grow.
The problem stems from the fact even with two continuous variables
that do not share any coordinate pairs, the size ratio between the
visual marks and the display remains more or less constant.
Furthermore, most datasets are not uniformly distributed.
This all means that overplotting is bound to happen for realistic
datasets.

Ellis and Dix~\cite{Ellis2007} survey the literature and derive a
general approach to reduce clutter.
According to their treatment, there are three ways to reduce clutter
in a visualization: by changing the visual appearance, through space
distortion, or by presenting the data over time.
Some trivial but impractical mechanisms they list include decreasing
mark size, increasing display space, or animating the data.
Below we review more practical approaches based on appearance and
distortion.

\subsection{Appearance-based Methods}

Practical appearance-based approaches to mitigate overplotting include
transparency, sampling, kernel density estimation (KDE), and
aggregation.
Transparency changes the opacity of the visual marks, and has been
shown to convey overlap for up to five occurrences~\cite{Zhai1996}.
However, there is still an upper limit for how much overlap is
perceptible to the user, and the blending caused by overlapping marks
of different colors makes identifying specific colors difficult.
Sampling uses stochastic methods to statistically reduce the data size
for visualization~\cite{Dix2002}.
This may reduce the amount of overplotting, but since the sampling
must be random, it can never reliably eliminate it.

KDE~\cite{Silverman1986} and other binned aggregation
methods~\cite{Elmqvist2010, Fua1999} replace a cluster of marks with a
single entity that has a distinct visual representation.
However, these methods are difficult to apply for scatterplots because
scatterplots operate on the principle of object identity, meaning that
each visual mark is supposed to represent a single entity.
Splatterplots~\cite{Mayorga2013} overcome this by overlaying
individual marks side-by-side with the aggregated entities, using
marks to show outliers and aggregated entities to show the general
trends.
However, even with only few aggregated entities, the resulting
color-blended image becomes visually complex and challenging to read
and understand.

\subsection{Distortion-based Methods}

Unfortunately, appearance-based clutter reduction
methods~\cite{Ellis2007} are not well-suited for discrete variables,
since such dimensions may cause many data points to map to the exact
same screen location.
In such a situation, changing the appearance of the marks does not
help.
For such data, distortion-based techniques may be better.
The canonical distortion technique is jittering, where a random
displacement is used to subtly modify the exact screen space position
of a data point.
This has the effect of spreading data points apart so that they are
easier to distinguish.
However, most na{\"i}ve jittering mechanisms apply the displacement
indiscriminately to all data points, regardless of whether they are
overlapping or not.
This has the drawback of distorting all points away from their true
location on the visual canvas, and still does not completely eliminate
overplotting.

Bezerianos et al.~\cite{Bezerianos2010} use a more structured approach
to displacement, where overlapping marks are organized onto the
perimeter of a circle.
The circle is grown to a radius where all marks fit, which means that
its size is also an indication of the number of participating points.
However, this mechanism still introduces uncertainty in the spatial
mapping, and it is also not clear how well it scales for very dense
data.
Nevertheless, it is a good example of how deterministic displacement
can be used to great effect for eliminating overplotting.

Trutschl et al.~\cite{Trutschl2003} propose a deterministic
displacement (``smart jittering'') that adds meaning to the location
of jittering based on clustering results.
Similarly, Shneiderman et al.~\cite{Shneiderman2000} propose a related
structured displacement approach called \textit{hieraxes}, which
combines hierarchical browsing with two-dimensional scatterplots.
In hieraxes, a two-dimensional visual space is subdivided into
rectangular segments for different categories in the data, and points
are then coalesced into stacked groups inside the different segments.
This idea is obviously very similar to our gatherplots technique, but
the main difference is that we in this paper derive gatherplots as
generalizations of scatterplots, and also define mechanisms for laying
out the stacked groups, organizing them by another dimension, and
modifying their aspect ratio to support relative assessments.

\subsection{Visualizing Nominal Variables}

While we have already ascertained that scatterplots are not optimal
for nominal variables, there exists a multitude of visualization
techniques that are~\cite{Bederson2002, Hofmann2000, Kosara2006}.
Simplest among them are histograms, which allows for visualizing the
item count for each nominal value~\cite{Stevens1946}, but much more
complex representation are possible.
One particular usage for visualizing nominal data that is of practical
interest is for making inferences based on statistical and
probabilistic data.
Cosmides and Toody~\cite{Cosmides1996} used frequency grids as
discrete countable objects, and Micallef et al.~\cite{Micallef2012}
build extend this with six different area-proportional representations
of nominal data organized into different classes.

As a parallel to our work on gatherplots, one particular
multidimensional visualization technique that is closely related to
scatterplots is parallel coordinate plots~\cite{Inselberg1985}.
However, just like scatterplots, parallel coordinate plots are often
plagued by overplotting due to high data density and discrete data
dimensions.
The work by Kosara et al.~\cite{Kosara2006} to extend parallel
coordinate plots into \textit{parallel sets} is interesting because it
specifically addresses the overplotting concern by grouping points
with the same value into a segment on the parallel axis.
This is precisely the same idea we will apply for scatterplots in this
work.

\section{The Gather Transformation}
\label{sec:gather}

Position along a common scale is the most salient of all visual
variables~\cite{Bertin1983, Cleveland1985}, and so mapping a data
dimension to positions on a graphical axis is a standard operation in
data visualization.
We call this mapping a \textit{visual transformation}.
However, most statistical treatments of data, such as Stevens'
classical theory on the scale of measurements~\cite{Stevens1946}, do
not take the physical properties of display space into account.
This is our purpose in the following section.

\subsection{Problem Definition}

Let $V = <f, s>$ be a visual transformation that consists of a
transformation function $f$ and a mark size $s$.
Furthermore, assume that $f$ transforms a data point $p_d \in D$ from
a data dimension $D$ to a coordinate on a graphical axis $p_c \in C$
by $f(p_d) = p_c$.
Given a dataset $D_i \subseteq D$, we say that a particular visual
transformation $V_j$ exhibits \textit{overlap} if

$$
\exists p_x, p_y \in D_i \wedge x \neq y : |f_j(p_x) - f_j(p_y)| < s_j.
$$

In other words, overlap occurs for a particular dataset and visual
transformation if there exists at least one case where the visual
marks of two separate data points in the dataset fall within the same
interval on the graphical axis.
The \textit{overlap index} of a dataset and visual transformation is
defined as the number of unique pairs of points that overlap.
For a one-dimensional visualization, only a single transformation is
used and the visualization and dataset is said to exhibit
\textit{overplotting} iff it exhibits overlap.
For a two-dimensional visualization, however, the visualization and
dataset will only exhibit overplotting iff there is overlap in
\textbf{both} visual transformations and data dimensions.
Analogously, the \textit{overplotting index} is the unique number of
overplotting incidences for that particular visual transformation and
dataset.

This has two practical implications: (1) even a dataset that consists
only of nominal variables may \textbf{not} exhibit overplotting if
there is only at most one instance of each nominal value, and (2) a
dataset consisting of continuous values may \textbf{still} exhibit
overplotting if any two points in the dataset are close enough that
they get mapped to within the size of the visual marks on the screen.
The corollary is basically that overplotting is a function of
\textbf{both} visualization technique and dataset.

\begin{figure*}[htb]
  \centering
  \includegraphics[width=7in]{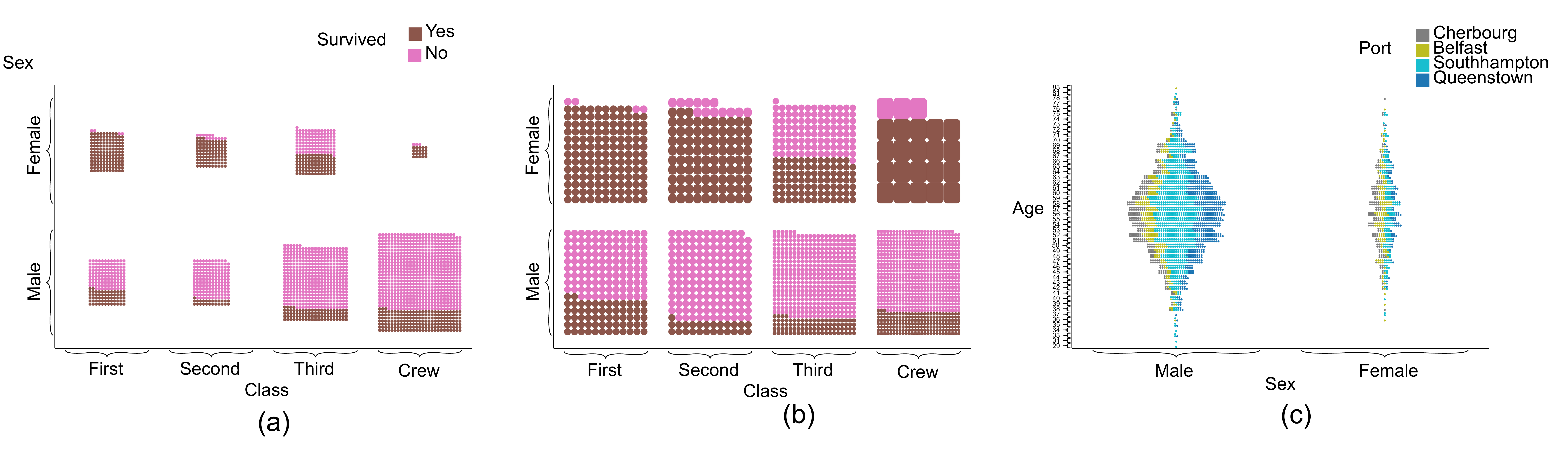}
  \caption{Main layout modes for gatherplots: (a) absolute mode with
    constant aspect ratio, which arranges items following the aspect
    ratio of given area; (b) relative mode of (a).
    The rate of survivors in each male passenger class is not each to
    compare.
    Figure (c) shows the streamgraph mode, where each cluster
    maintains the number of element in the shorter edge, making it
    easier to see the distribution of the subgroups along the Y axis.}
  \label{fig:aspectRatio}
\end{figure*}

\subsection{Definition: The Gather Transformation}

We build on the previous idea of structured
displacement~\cite{Bezerianos2010, Shneiderman2000} by proposing a
novel visual transformation function called a \textit{gather}
transformation $f_{gather}$ that non-linearly segments the graphical
axis $C$ and organizes data points in each segment to eliminate
overplotting.

The gather transformation $V_{gather} = <f_{gather}, s_{gather}>$
consists of a transformation function $f_{gather}$ that maps data
points $p_d \in D$ to coordinates $p_c \in C$, and a visual mark
sizing function (instead of a scalar) $s_{gather}$ that yields a
visual mark size given the same data point.
The gather transformation function is special in that it eliminates
overplotting by subdividing the graphical axis $C$ into $n$ contiguous
segments $C = \{ C_1, C_2, \ldots, C_n \}$, where $n$ is the size of
the domain of the gather transformation function, i.e., the number of
unique elements in the data dimension $D$.
When mapping a data point $p_d$ to the graphical axis, $f_{gather}$ will
return an arbitrary graphical coordinate $p_c \in C_i$ for whatever
coordinate segment $C_i$ that $p_d$ belongs to.

Practically speaking, coordinates $p_c \in C_i$ will be chosen to
efficiently pack visual marks into the available display space without
causing overplotting (i.e., using a regular spacing of size
$s_{gather}$).
Several different methods exist for adapting the gather transformation
to the dataset $D$.
One approach is to keep the segments $C_1, \ldots, C_n$ of equal size
and find a constant visual mark size $s_{gather}(p_d) = s_{max}$ that
ensures that all points fit within the most dense segment.
The constant mark size makes visual comparison straightforward.
Another approach is to adapt segment size to the density of the data
while still keeping the mark size constant.
This will minimize empty space in the visual transformation and allows
for maximizing mark size.
A third approach is to vary mark size proportionally to the number of
points in a segment.
This will make comparison of the absolute number of points in each
segment difficult, but may facilitate relative comparisons if marks
are distinguished in some other way (e.g., using color).

For data dimensions $D$ that have a very large number of unique
values, it often makes sense to first quantize the data using a
function $p_q = Q(p_d)$ so that the number of elements $n$ is kept
manageable (on the order of 10 or less for most visualizations).
For example, a data dimension representing a person's age might
heuristically be quantized into ranges of 10 years: 0-9 years, 10-19
years, 20-29 years, and so on.

In a gather transformation, the coordinate axis has been partitioned
into segments, where the order of segments on the axis depends on the
data.
For nominal data, the segments can be reordered freely, both by the
algorithm and by the user.
For ordinal or quantized data, the order is given by the data
relation.
Furthermore, it often makes sense to be able to order points inside
each segment $C_i$ using the gathering transformation function
$f_{gather}$, for example using a second data dimension (possibly
visualized using color) to group related items together.

Appropriate visual representations of data where the gather
transformation has been applied are also important.
The \textit{stacked entities} of gathered points---one per coordinate
segment $C_i$---should typically maintain object identity, so that
each constituent point and their size is discernible as a discrete
visual mark.
Similarly, a visual representation of the segmented graphical axis
should externalize the segments as labeled intervals instead of
labeled major and minor ticks; this will also communicate the
discontinuous nature of the axis itself to the viewer.

\subsection{Using the Gather Transformation}

To give an example in one-dimensional space, parallel coordinate
plots~\cite{Inselberg1985} use multiple graphical axes, one per
dimension $D_i$, and organize them in parallel while rendering data
points as polylines connecting data values on one axis to adjacent
ones.
However, traditional parallel coordinate plots merely use a scatter
transformation on each graphical axis, which makes the technique prone
to overplotting.
Multiple authors have studied ways of mitigating this problem, for
example by reorganizing the position of nominal
values~\cite{Rosario2004}, using transparency, applying jitter, or by
clustering the data~\cite{Fua1999}.

However, an alternative approach is to use the gather representation
for each graphical axis to minimize overplotting.
This will cause each axis to be segmented into intervals, and we can
then resize segments according to the number of items falling into
each segment so that segments with many data points become
proportionally larger than those with fewer points.
Finally, if the data dimensions represent nominal data, it may make
sense to use a global segment ordering function so that there is a
minimum of lateral movement for the majority of points as they
connect to adjacent axes.
This will also minimize line crossings between the parallel axes.
This particular visualization technique---a parallel coordinate plot
with the gather transformation applied to each graphical axis---is
essentially equivalent to \textit{parallel sets}~\cite{Kosara2006}.

In fact, by applying our generalized gather transformation to the
axis, we are actually proposing a new type of stacked visualization
where each entity is still represented by lines.
In a sense, this technique combines parallel coordinates and parallel
sets because the grouped lines maintain the illusion of a single
entity for an axis with nominal categorical values (similar to
parallel sets), yet integrates directly with a parallel coordinate
axis with continuous values.
The main difference is that the new parallel coordinate/set variation
allows each axis to be either categorical or continuous, meaning that
one axis can represent the gender and the next can represent the
height of person.

\section{Gatherplots: A 2D Gathering Representation}

Here we apply the concept of gathering to the scatterplots to
alleviate overplotting, focusing on optimal layouts of gathered
entities, graphical representations of chart elements, and novel
interactions.

\subsection{Applying Gathering to Perpendicular Axes}

The application of the gather transformation results in the
segmentation of output range of an axis scale, where the items with
same values will be arranged to avoid overplotting.
Applying gathering to two perpendicular axes defining a Cartesian
space results in a \textit{gatherplot}: a 2D visualization technique
that aggregates entities for each axis.
However, given this basic visual representation, there are many open
design possibilities for aspect ratio, layout, and item shapes.
We discuss these design parameters in the treatment below.

\subsection{Layout Modes}

Gatherplots organizes entities into \textit{stacked groups} according
to a discrete variable to eliminate overplotting.
However, the result depends on the context, especially on the size
distribution of each groups, the aspect ratio of assigned space, and
the task at hand.
This makes finding an optimal layout difficult.
One solution is to provide interaction techniques to change the
layout, but such layout options may lead to confusion for the user.
Our approach is to provide one general optimal visualization for the
most common aspect ratio and tasks, and provide a very few optional
methods to change it.
As a result, we derive the following three layout modes (examples in
Figure~\ref{fig:aspectRatio}):

\begin{itemize}

\item\textbf{Absolute mode.} Here stacked groups are sized to follow the
  aspect-ratio of the assigned region.
  The node size of the items are determined by the maximum length dots
  which can fill the assigned region without overlapping.
  This means with the same assigned space, the groups with the maximum
  number of members determines the overall size of the nodes.

\item\textbf{Relative mode.} In this mode, the node size and aspect
  ratio is adapted so that every stacked group has equal dimensions.
  This is a special mode to make it easier to investigate ratios when
  the user is interested in the relative distributions of subgroups
  rather than the absolute number of members.
  Items also change their shape from a circle (absolute mode) to a
  rectangle.

\item\textbf{Streamgraph mode.}  Here stacked groups are reorganized
  so that the maintain the same number of elements in their shorter
  edge.
  This mode is used for regions where the ratio of width and height
  are drastically different (in our prototype implementation, we use a
  heuristic value of 3 for aspect ratio to be a threshold for
  activating this mode).
  This means there are usually many times more groups in the axis in
  parallel with shorter edges.
  The purpose of this mode is to make it easier to compare the size
  along these many entities.
  The resulting graphic resembles ThemeRiver~\cite{havre2000} as the
  number of entities increase.

\end{itemize}

The choice between absolute mode and streamgraph mode happens
automatically based on the aspect ratio and without user intervention.
Therefore only interactive option is required to toggle between
absolute mode and relative mode.
Our intuition is that the absolute mode should be good enough for most
of the time, and when very specific tasks are required, the user can
switch to the relative modes.

However, gatherplots involve many more possibilities beyond than these
layout functions.
Below follows our treatment of these design possibilities and our
rationale for our decisions.

\subsubsection{Area vs. Length Oriented Layout}

\begin{figure}[htb]
  \centering
  \includegraphics[width=3in]{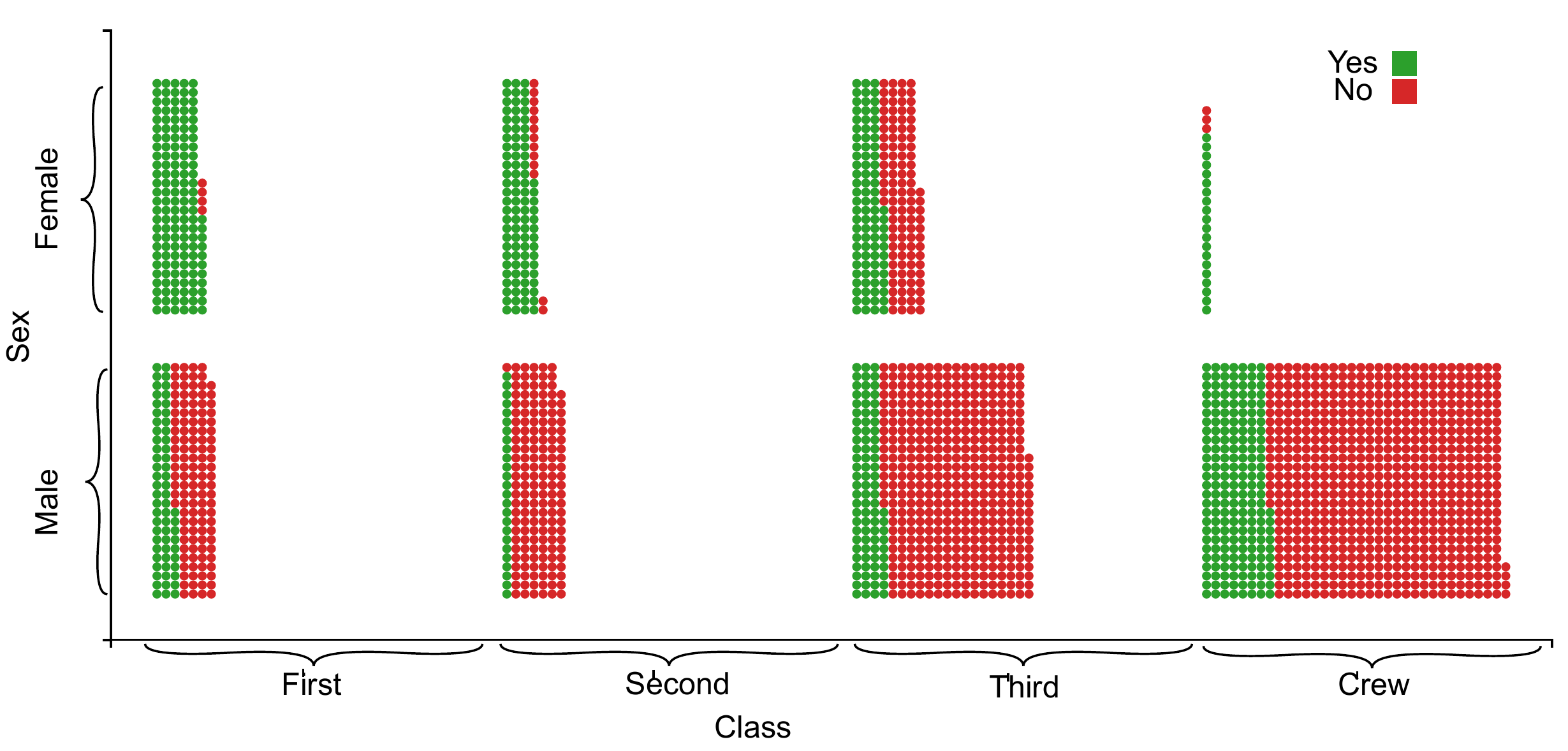}
  \caption{Stacked group layouts for gathering.
    This layout supports comparison group sizes.
    Because the height of stacked groups is all fixed to the same
    value, comparing the length yields the size.}
  \label{fig:lengthBasedLayout}
\end{figure}

Maintaining the aspect ratio of all stacked groups means that the size
of the group is represented by its area.
The length of the group is only used in special cases when the aspect
ratio is very high or low.
According to Cleveland and McGill~\cite{Cleveland1985}, length is far
more effective than the area for graphical perception.
However, Figure~\ref{fig:lengthBasedLayout} shows the three problems
associated with layout to enable length-based size comparison.
In this view, the items are stacked along the vertical axis to make
the size comparison along the horizontal axis easier.
The width of rectangle is all set to be equal to so that the length
can represent the size of subgroups.
However, they show drastically different shape of line vs.\ rectangle,
which may cause users to lose concept of equality.
Furthermore, to make length-based comparison easier, the stacking
should be aligned to one side of the available space: left, right,
top, or bottom.

In this case, the bottom is selected to make it easier to compare
along the X axis.
However, this creates additional two problem.
The first problem is that center of mass of each stacked group is so
different that the concept of belonging to the same value can be
misleading.
The second problem is that choosing alignment direction is arbitrary
and depends on the task.
For example, in this view it is more difficult to compare along the Y
axis.
In this sense, this layout is biased to the X axis, while sacrificing
the performance along the Y axis.
For this reason, the most general choice is to use center alignment
with aspect ratio resembling the assigned range to avoid bias.

\subsubsection{Uniform vs. Variable Area Allocation}

\begin{figure}[htb]
  \centering
  \includegraphics[width=3in]{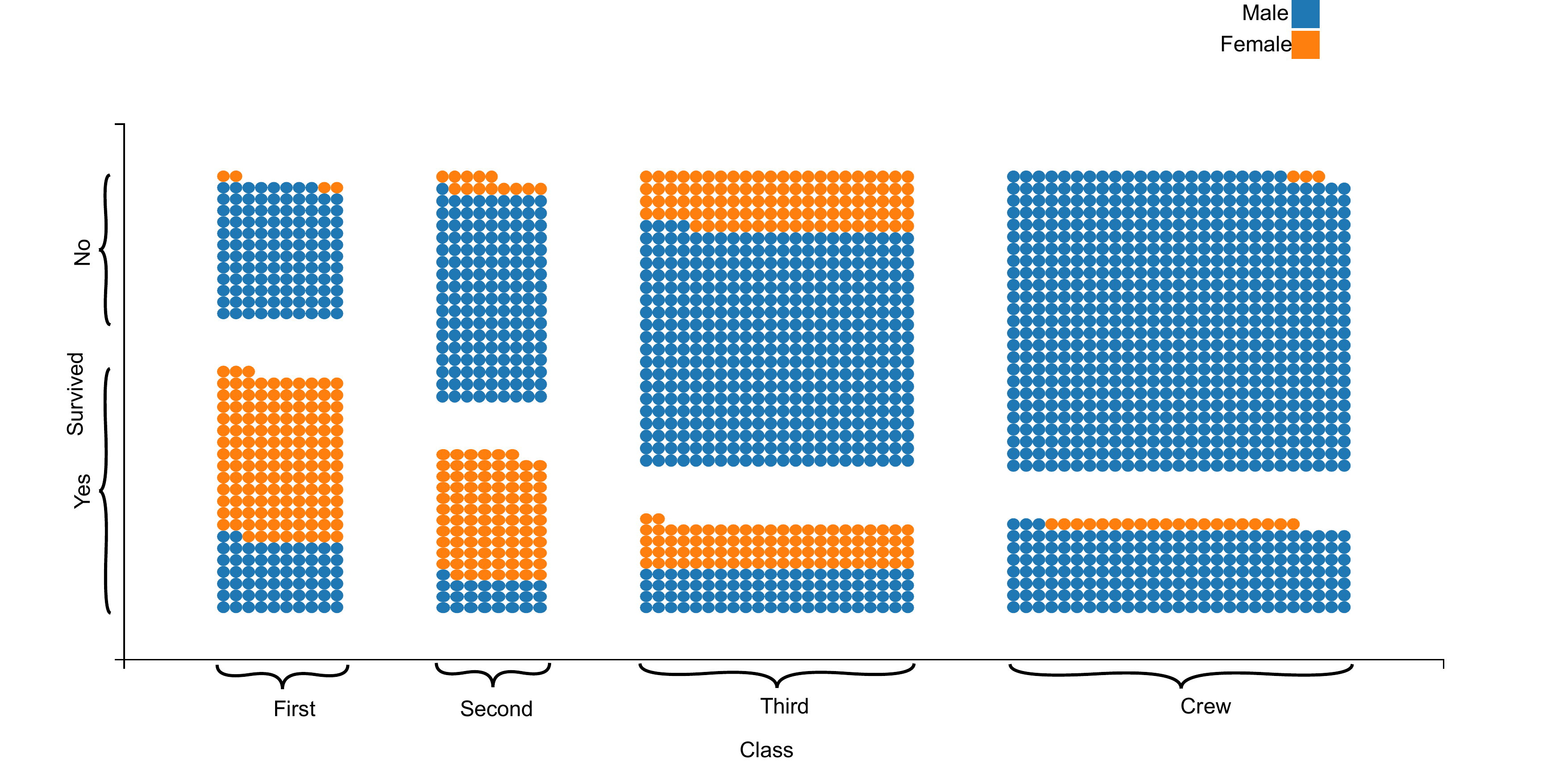}
  \caption{Variable area for a gatherplot.
    The chart uses space efficiently.
    Items which belong to same place should belong to the same value.
    However, as can be shown in the Y axis tick marker, this results
    in a violation of the scatterplot concept.}
  \label{fig:mosaic}
\end{figure}

In gatherplots, we assign uniform range to different values of
overlappable variables.
For some cases, assigning variable area can make sense and create
interesting visualizations.
As a simple example, we can argue that assigning the range of output
for gather transfer function to be proportional to the numbers of
items that belong that value uses the space most efficiently.
This will result in the following layout shown in
Figure~\ref{fig:mosaic}, which is basically a mosaic
plot~\cite{hartigan1981}.

\subsubsection {Role of Relative Mode}

Since gathering assigns a discrete noncontinuous range to each
graphical axis, each stacked group can be grown to fill all available
space.
This relative mode is useful for two specific tasks:

\begin{itemize}
\item Getting a relative percentage of the subgroups in the group
  (Figure~\ref{fig:aspectRatio}).
  Because groups of different size is normalized to the same size, any
  comparison in area results in a relative comparison.
\item Finding the distribution of outliers.
  When there are many items on the screen for absolute mode, all node
  sizes must be reduced.
  This can make outliers hard to locate.
\end{itemize}

\subsection {Graphical Representations}

In this section, we discuss the visual representation for gatherplots
and how it differs from traditional scatterplots.

\subsubsection{Continuous Color Dimension}

\begin{figure}[htb]
  \centering
  \includegraphics[width=3in]{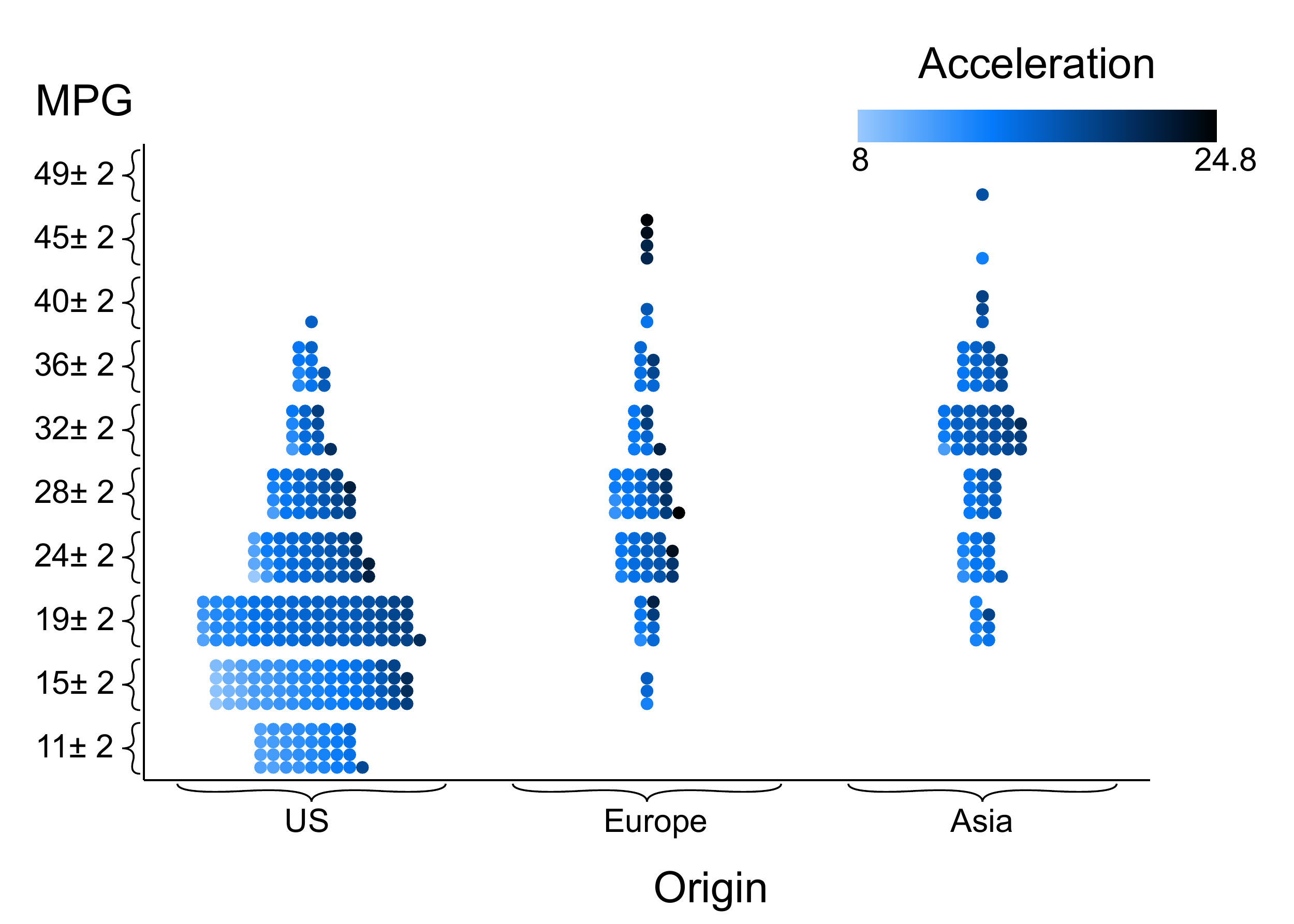}
  \caption{Continuous color scale used in a gatherplot.
    The X axis is origins of cars and the Y axis is MPG.
    The color scale is acceleration.}
  \label{fig:continuousColor}
\end{figure}

The gather transform sorts items according to a data property, such as
a variable also assigned for coloring items.
This removes the scattered color patterns in the stacked groups that
is common in other techniques such as Gridl~\cite{Shneiderman2000}.
This is also particularly useful for continuous color scales, making
the variation of colors are easier to perceive
(Figure~\ref{fig:continuousColor}).

\subsubsection{Shape for Items}

\begin{figure}[htb]
  \centering
  \includegraphics[width=3in]{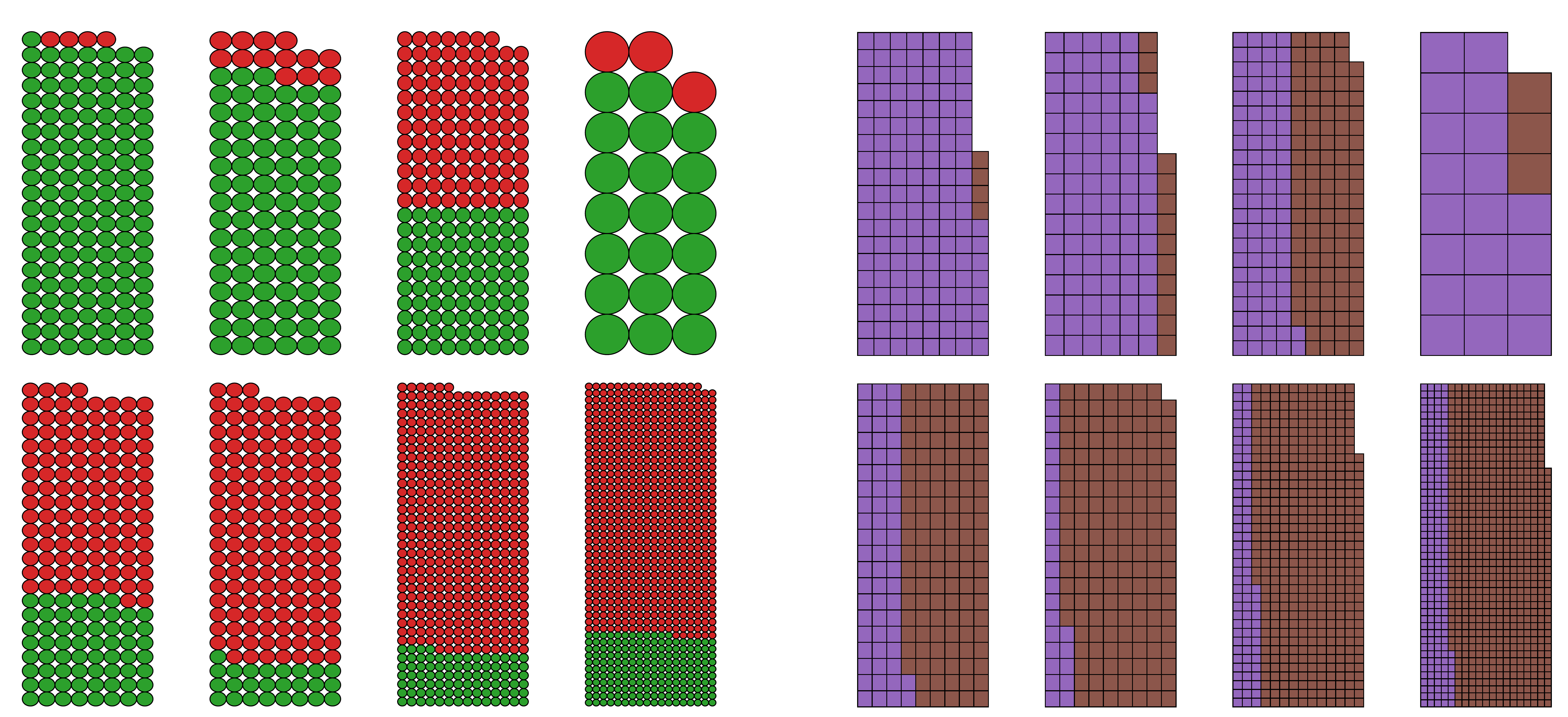}
  \caption{Stroke line problems where the circle consumes ample white
    space between adjacent nodes, which contributes to clutter as it
    grows.
    The rectangle does not have space between nodes, however, it must
    have a stroke border to show stretching.
    But this borderline creates problems when the items are very
    small.}
  \label{fig:shapes}
\end{figure}

Scatterplots typically use a small circle or dot as a visual
representation for items, but many variations exist that use glyph
shapes to convey multidimensional variables\cite{Mcdonnel2009,
  Tufte1983, carr1983interactive, Cleveland1988, chernoff1973use}.
However, in the relative mode, sometimes the aspect ratio of nodes
changes according to the aspect ratio of box assigned to that value.
Also, as gathering changes the size of nodes to fit in one cluster,
sometimes node size becomes too small, or too large compared to other
nodes.
This results in several unique design considerations for item shapes.
After trying various design alternatives, we recommend using a
rectangle with constant rounded edge without using stroke lines.
Using constant rounded edge allows the nodes to be circular when the
node is small, as in Figure~\ref{fig:aspectRatio}(b), and a rectangle
to show the degree of stretching, as shown in
Figure~\ref{fig:aspectRatio}(b).
Figure~\ref{fig:shapes} shows some previous trials with various
shapes.

\subsubsection{Design of Tick Marks }

\begin{figure}[htb]
  \centering
  \includegraphics[width=3in]{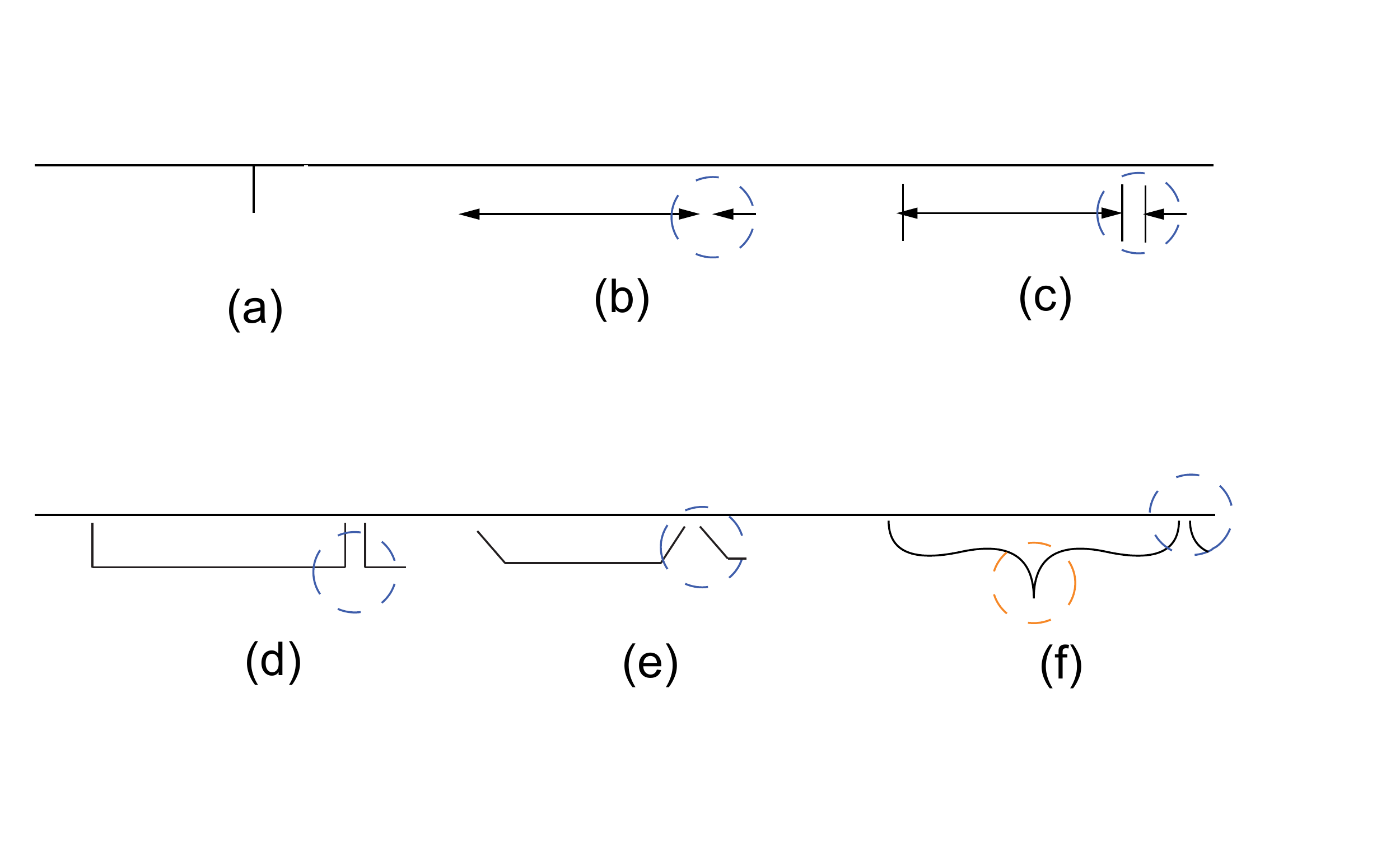}
  \caption{Various tick marks types.
    The blue dotted region represents the area between adjacent tick
    marks.
    (a) is a typical line type tick mark for the scatterplots.
    (b) lacks guide lines, which will make anchoring easier.
    (c) creates a packed region between adjacent marks.
    (d) uses less crowded region in this regionn, but (e) is the least
    crowded.
    (f) is the final recommendation, with the data label in the orange
    region.}
  \label{fig:tickMark}
\end{figure}

The single line type tick marks for scatterplots are not appropriate
for gatherplots.
Because we are representing a range rather than a single point, a
range tick marker will be better.
Without this visual representation, when the user is confronted with a
number, it can be confusing to determine whether adjacent nodes with
different offset has same value or not.
After considering a few visual representation, we recommend a bracket
type marker for this purpose.
Figure~\ref{fig:tickMark} shows various types of markers for range
representation.
The bracket is optimal in that it uses less ink and creates less
density with adjacent ticks.

\subsubsection{Tick Labels for Numbers}

\begin{figure}[htb]
  \centering
  \includegraphics[width=3in]{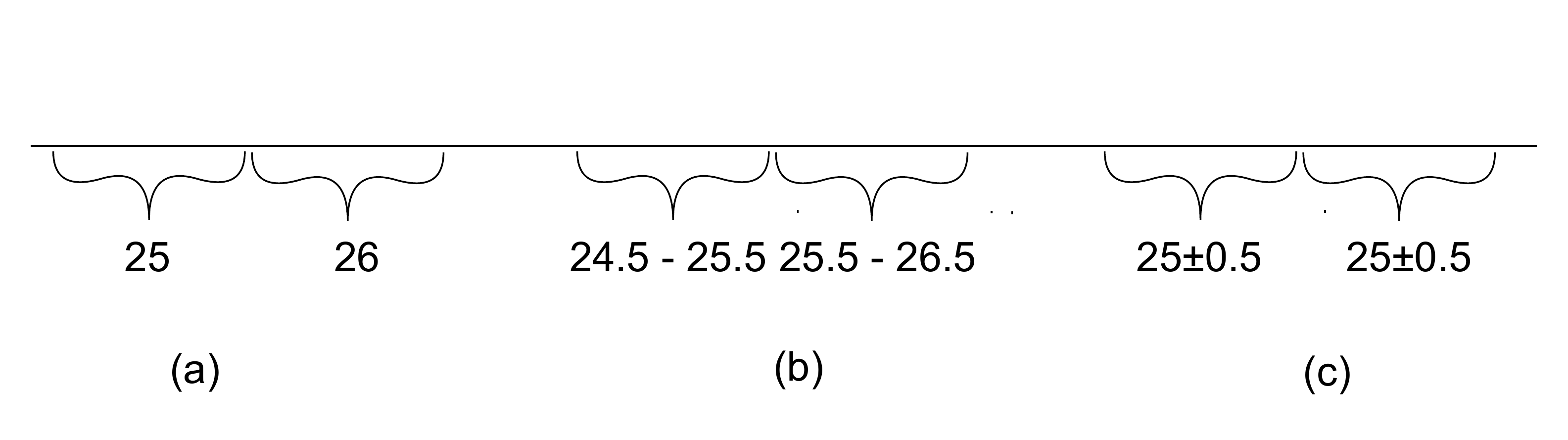}
  \caption{Several different tick label designs, including (a) numbers
    (used for nominal values), (b) extents of a range, and (c) the
    size of the range.}
  \label{fig:tickLabel}
\end{figure}

In gatherplots, a tick mark represents a range.
This creates a problem for the data label.
Because it is a number, the na{\"i}ve way to represent will be having
beginning number and ending number at each side, as shown in
Figure~\ref{fig:tickLabel}.
However this creates a very dense region between adjacent marks and,
worse, the same number is repeated in this region, thereby wasting
space.
We recommend the use of a plus-minus sign to represent a bin size, to
create a conceptually consistent tick labels.
One limitation with this approach is the binning of arbitrary size can
create bins of arbitrary floating number occupying an inappropriate
amount of size.

\subsubsection{Extension of Scatterplot Matrix}

\begin{figure}[htb]
  \centering
  \includegraphics[width=3in]{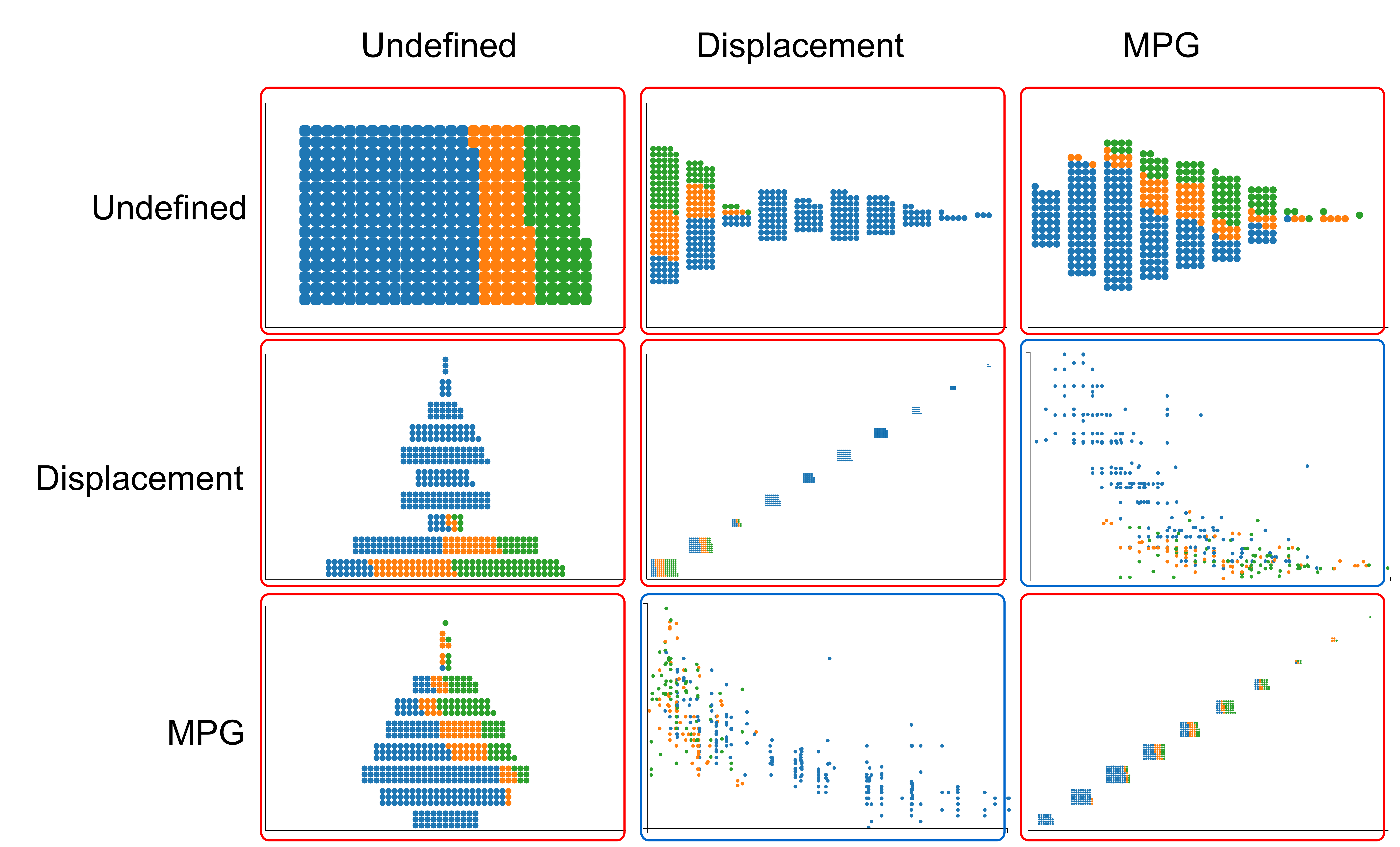}
  \caption{This shows the power of an \textit{undefined} axis using an
    example of a gatherplots matrix for the cars dataset.
    Two continuous variables, Displacement and MPG, are used.
    The charts in the red box are new addition to the previous
    scatterplots matrix, while the charts in the blue box remains the
    same.
    Using the undefined axis enabled creating an informative chart
    showing the distribution of only single variable with a
    streamgraph type graphic.
    The upper right corner of the matrix shows undefined
    vs.\ undefined, which is logically representing all the items.}
  \label{fig:matrix}
\end{figure}

One subtle difficulties in working with scatterplot matrix is when
the users want to see only single axis, without definition in other
axis.
Or one may assigning same dimension to the X and Y axes while
exploring various dimensions.
Both will create an overdrawing in a traditional scatterplot matrix.
However, in gatherplots, an undefined axis result in the aggregation
of all nodes in one group, which is a spontaneous logical extension.
This enables the scatterplots matrix to have an additional row and
column with undefined axes.
Figure~\ref{fig:matrix} shows an example of this using cars dataset.
Two dimensions---displacement and MPG---were used to create a 3 by 3
gatherplots matrix.
Note that 7 out of 9 charts are new compared to scatterplots, while
adding information to the whole picture.
The diagonals also enable seeing distribution, which is a improvement
over previous scatterplots matrix.

\subsubsection{Applications for Continuous Variables}

\begin{figure*}[htb]
  \centering
  \includegraphics[width=7.5in]{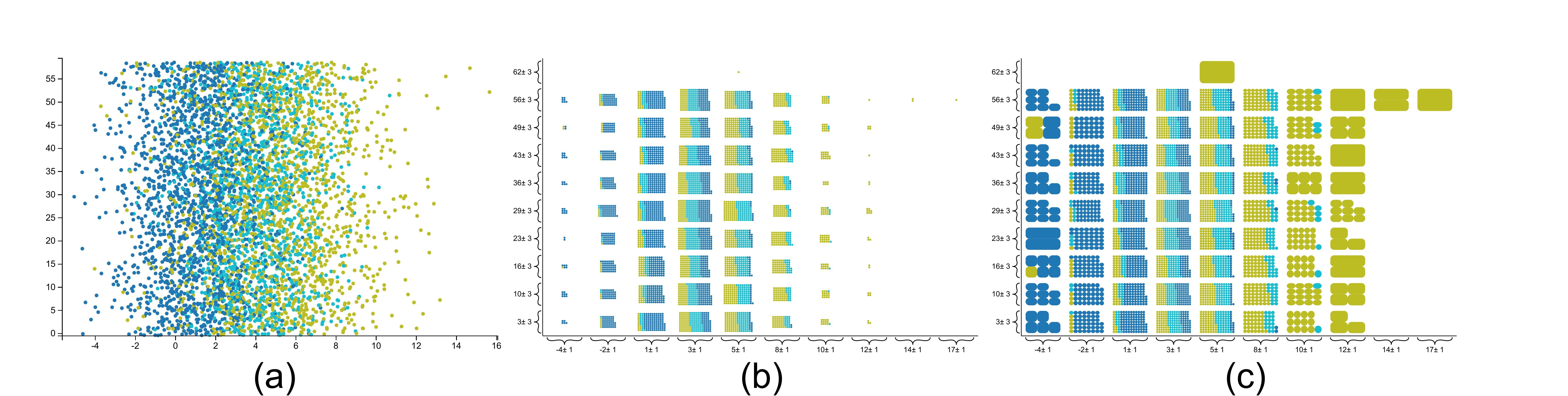}
  \caption{Using gatherplots to manage overplotting.
    (a) shows a scatterplot with 5,000 random numbers with severe
    overplotting in the center area.
    In (b), gathering is applied to create a more organized view.
    However, the gathering resizes the items so small that it becomes
    difficult to detect outliers.
    (c) shows relative mode, where the outliers are enlarged.
    This makes identifying the distribution of sparse regions easier.}
 \label{fig:contcont}
\end{figure*}

Gatherplots can be used to mitigate overplotting caused by
continuous variables as well.
Figure~\ref{fig:contcont} (a) shows how gatherplots handle the
overplotting caused by continuous variables.
The plot is using relative mode with two random variables.
The relative mode makes it easier to identify the outliers and the
distribution of outliers.

One limitation of gatherplots is that it requires binning to manage a
continuous variable, yet binning creates arbitrary boundaries.
In this sense, gatherplots can be misleading.
However, combining gatherplots with scatterplots makes this problem
less severe.

\subsubsection{Animated Transitions}

The many shape and layout transitions involved in gatherplots can be
confusing to users.
Animation can be a powerful tool to reduce this and maintain the
user's mental model.
Heer and Robertson investigated the effectiveness of animated
transitions and found that animating the statistical chart can improve
the perception of statistical data\cite{heer2007animated}.
Robertson et al.~\cite{robertson2008effectiveness} found that
animation leads to an enjoyable and exciting experience, even ifx the
analysis was not effective.
Elmqvist et al.~\cite{Elmqvist2008} used animation for the
scatterplots to maintain congruence.
Drawing on all of this work, gatherplots uses animated transitions for
all state and layout changes.
In addition to the animation according to the axis dimension change,
animation is used to show thx transition between scatterplots and
jitterplots.
This ameliorates the potential misconception of the data distribution
in the gatherplots.

\subsubsection {Axis Folding Interaction}

\begin{figure}[htb]
  \centering
  \includegraphics[width=3in]{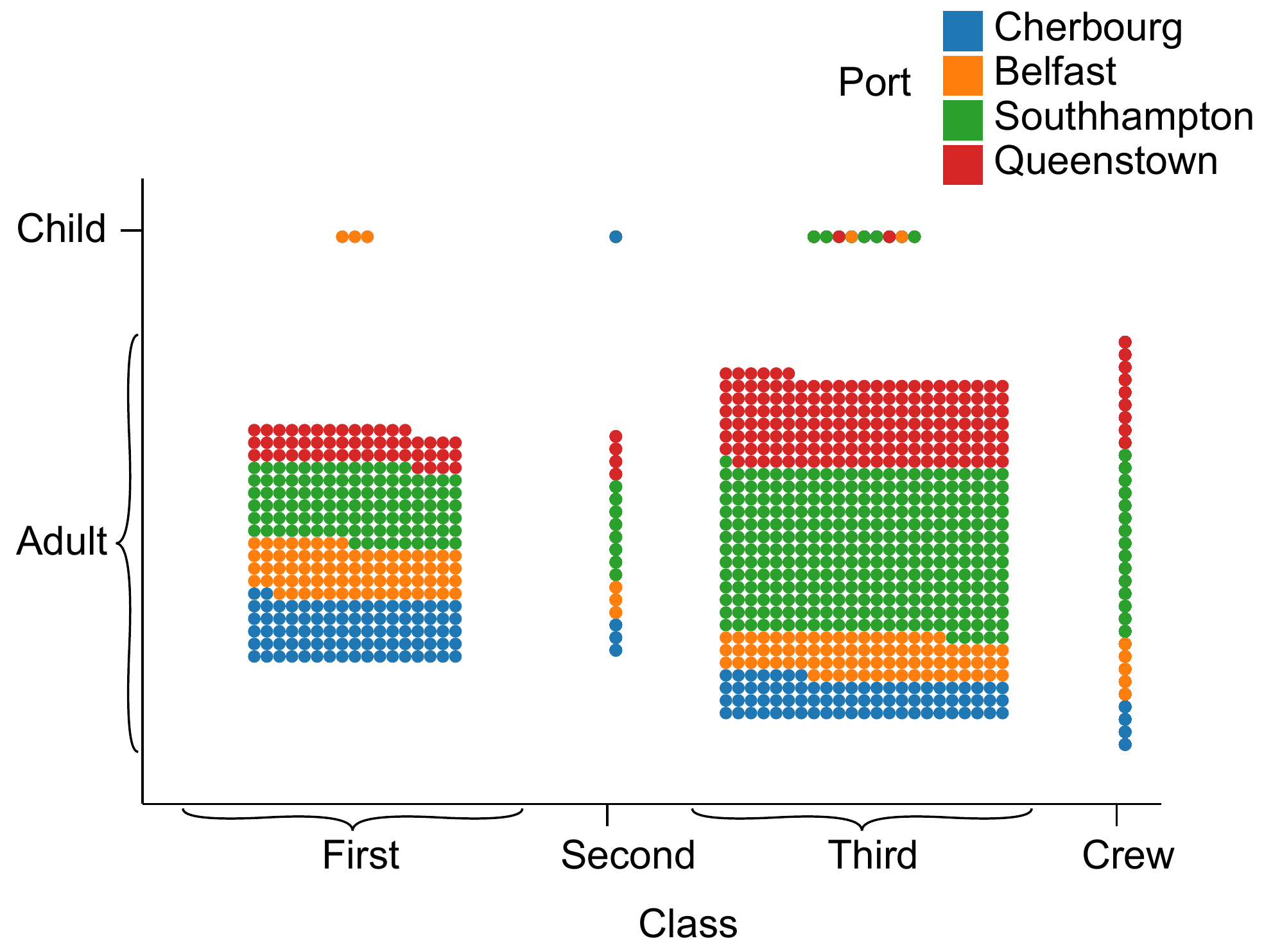}
  \caption{Survivors of the Titanic using a gatherplot.
    The X axis is class of passengers, where second class passengers
    and crew are minimized.
    The Y axis is age, where the adult value is maximized.
    This view makes it easy to compare first class adults and third
    class adults.
    Note that even in the minimized state, we can get an overview
    about the second class and crew by the color line, which
    communicates the underlying distribution.
    This is due to the sorting over the color dimension.}
  \label{fig:axisFolding}
\end{figure}

As an exploration tool for real-world dataset, it is crucial to have
means to filter unwanted data.
To aid this process with gather transform, we provide an optional
mechanism to go back to the original continuous linear scale function.
We allow each axis tick have an interactive control to be filtered out
(minimize) or focused (maximized).
This is called \textit{axis folding}, because it can be explained
mentally by a folding paper.
When minimized or folded, the visualization space is shrunk by
applying linear scales instead of nonlinear gather scales.
This results in overplotting as if a scatterplot was used for that axis.
A maximization is simply folding all other values except the value of
the interest in order to assign maximum visual space to that value.
Figure~\ref{fig:axisFolding} shows the axis folding applied to third
class adult passengers in the Titanic dataset.

\begin{figure}[htb]
  \centering
  \includegraphics[width=3in]{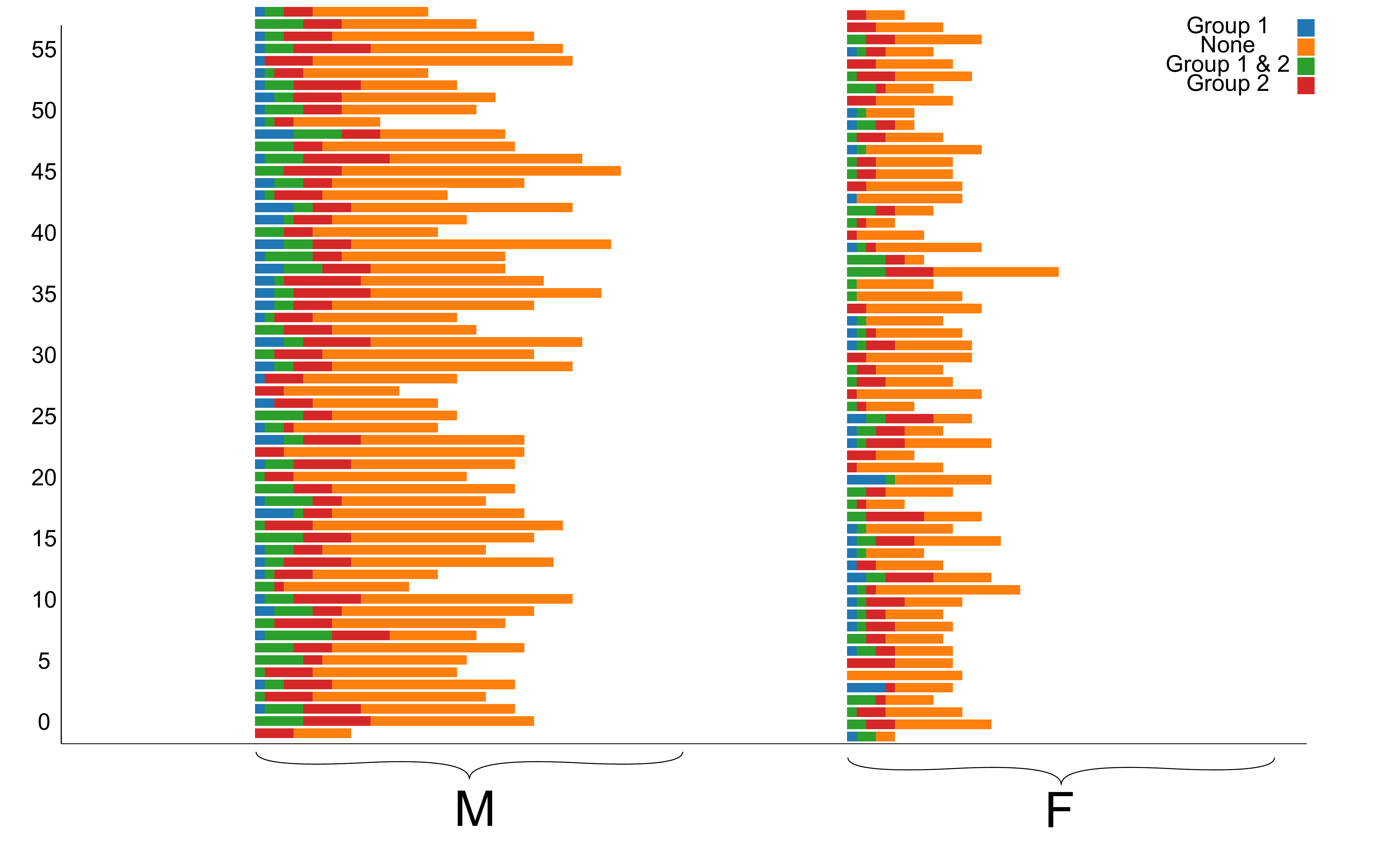}
  \caption{Hypothetical dataset with gender and age.
    Left alignment makes the distribution along the vertical direction
    more feasible.
    Note that only the vales in most left are anchored, so dimensional
    reordering is required.}
 \label{fig:genderVSAge}
\end{figure}

\section{Implementation}

We have implemented a web-based demonstration of
gatherplots and published it online.\footnote{\url{http://www.gatherplot.org}}
The users can load various dataset and compare each visualization with
scatterplots and jittered scatterplots with one button click.
In the top right area, an interactive guided walk-through is provided.
The users can follow instructions step-by-step to experience
gatherplots.
In the bottom, a discussion board is provided.
The purpose of the discussion board is to accumulate discussions
during the expert review.
Other people can also join the evaluation process.

\begin{figure}[htb]
 \centering
 \includegraphics[width=3.5in]{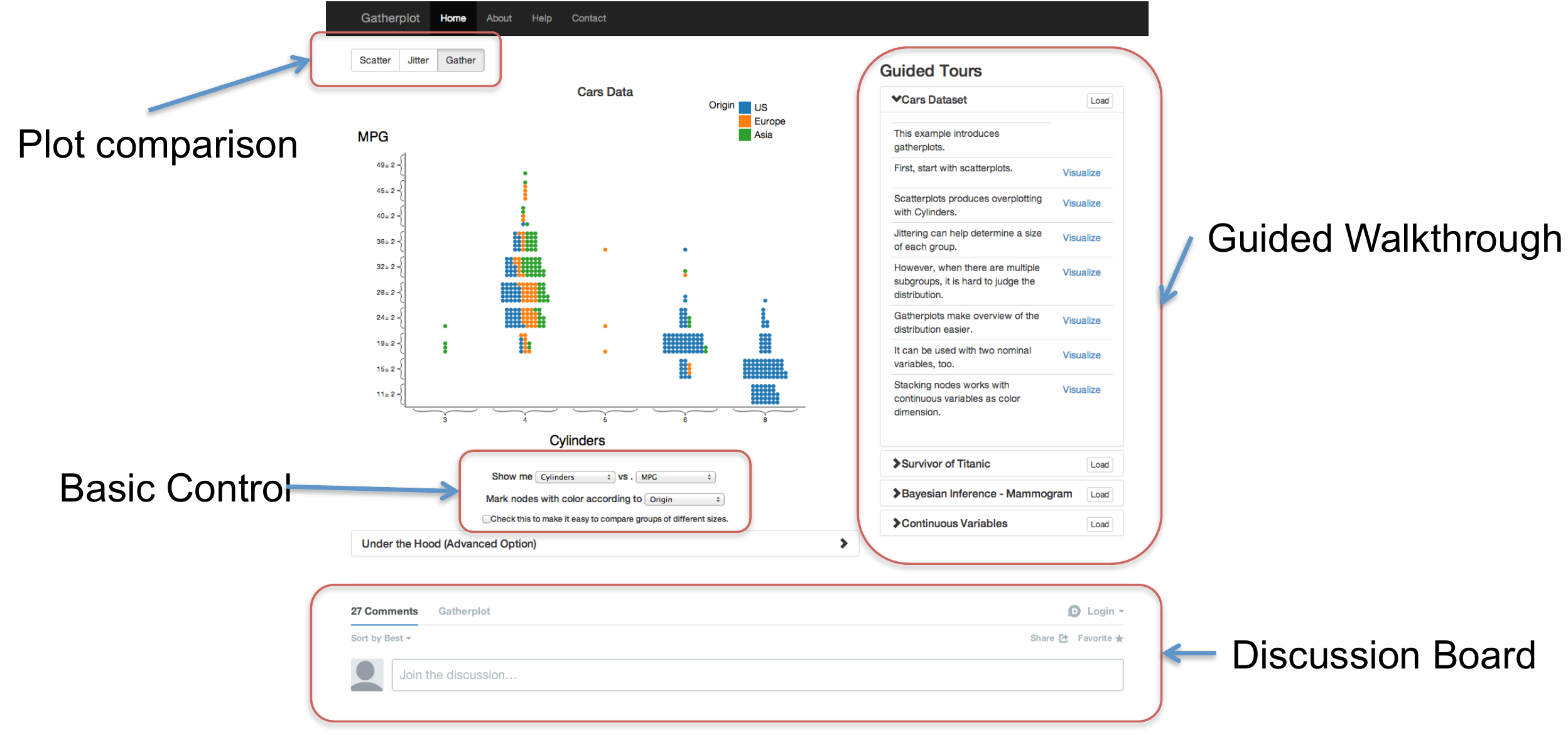}
 \caption{Web-based prototype implementation of gatherplots.
   Important features are comparison of gatherplots with scatterplots
   and jittered scatterplots, a guided walkthrough, and a discussion
   board.}
 \label{fig:screenshot}
\end{figure}

The gatherplot prototype implementation was developed using
D3.js\footnote{\url{http://www.d3js.org}} and
Angular.js\footnote{\url{http://www.angularjs.org}}.
Figure ~\ref{fig:screenshot} shows the screenshot of the implemented
website.
To test various layout and shape of the nodes, an intermediate version
which allows various tweak is also
available.\footnote{\url{http://www.gatherplot.org}}

\section{Evaluation of Gatherplots using Crowdsourcing}

The purpose of this study is to examine the effectiveness of gatherplots especially to see how different modes of gatherplots influence certain types of tasks for the crowdsourced workers.
Crowdsourcing platforms have been widely used and have shown to be reliable platforms for evaluation studies~\cite{paolacci2010running,willett2013identifying}.
Therefore, we conducted our experiment on Amazon Mechanical Turk~\footnote{\url{https://www.mturk.com}} .


\subsection{Experiment Design}

Gatherplots was developed to overcome limitations of conventional scatterplots.
We believe that gatherplots solves the issue of overdrawing, while maintaining structural identity with scatterplots.
Jittered scatterplots were selected as a comparison, as it is widely accepted standard technique maintaining same consistency with scatterplots.
We also wanted to measure how different modes of gatherplots were effective.
Therefore we designed the experiment to have four conditions such as scatterplots with jittering (jitter), gatherplots with absolute mode (absolute), gatherplots with relative mode (relative), and gatherplots with one check button to switch between absolute and relative mode (both).
We adopted between-subject design to eliminate learning effect by experiencing other modes.
The exact test environment is available for review~\footnote{\url{https://purdue.qualtrics.com/SE/?SID=SV_9YX7LCgsiwv0Voh}}. Note the questions for each conditions were generated randomly.

\subsection{Participants}

A total of 240 participants (103 female) completed our survey.
Because some questions asked a concept of absolute numbers and probability, we limited demographic to be United States to remove the influence of language.  Also to ensure the quality of the workers, qualification of workers were the approval rate of more than 0.95 with number of hits approved to be more than 1,000.
Only three of 240 participants did not use English as their first language. 119 people had more than bachelor's degree, with 42 people having hight school degree.
We filtered random clickers, if the time to complete one of questions was shorter than a reasonable time, 5 seconds.  Eventually, we have a total of 211 participants.

\subsection {Task}

Different layouts of gatherplots could support different types of tasks.
After reviewing task for nominal variables, we selected three types of task such as retrieving value as a low-level task; comparing and ranking as a high-level task.
For the comparing and ranking task, two different types of questions were asked: the tasks to consider absolute values such as frequency and tasks that consider relative values such as percentage.
Therefore, for one visualization 5 different questions were generated.
For gatherplots, our interest is more about the difference between questions considering absolute values and relative values.
The five types of questions are as follows:

\begin{itemize}
\setlength{\itemsep}{0.05pt}
\item\textbf{Type 1:} retrieve value considering one subgroup
\item\textbf{Type 2:} comparing of absolute size of subgroup between groups
\item\textbf{Type 3:} ranking of absolute size of subgroup between groups
\item\textbf{Type 4:} comparing relative size of subgroup between groups
\item\textbf{Type 5:} ranking relative size of subgroup between groups
\end{itemize}

To reduce the chance of one chart being optimal by luck for specific task, two charts of same problem structure were provided.  Eventually, the resulting questions were 10 for each participant.  Each question was followed by the question asking confidence of estimation with a 7-point Likert scale, and the time spent for each question was measured.

%

\subsection{Hypotheses}

We believe that different types of tasks will favor from different type of layouts.
Therefore our hypotheses are as follows:

\begin{itemize}
\setlength{\itemsep}{0.05pt}
\item[H1] For retrieving value considering one subgroup (Type 1), absolute, relative, both mode reduces the occurrence of the error than jitter mode.
\item[H2] For tasks considering absolute values (Type 2 and 3), the absolute mode reduces the error.
\item[H3] For tasks considering relative values (Type 4 and 5), the relative mode reduces the error.
\end{itemize}

\subsection{Results}

The results were analyzed with respect to the accuracy (correct or incorrect), time spent, and confidence of estimation.
Based on our hypotheses, we analyzed the different modes for each type of question: retrieve value, absolute value task, and relative value task.

\subsubsection{Accuracy}

\begin{figure}[h!]
\centering
\subfigure[]{
    {\includegraphics[width=0.4\columnwidth]{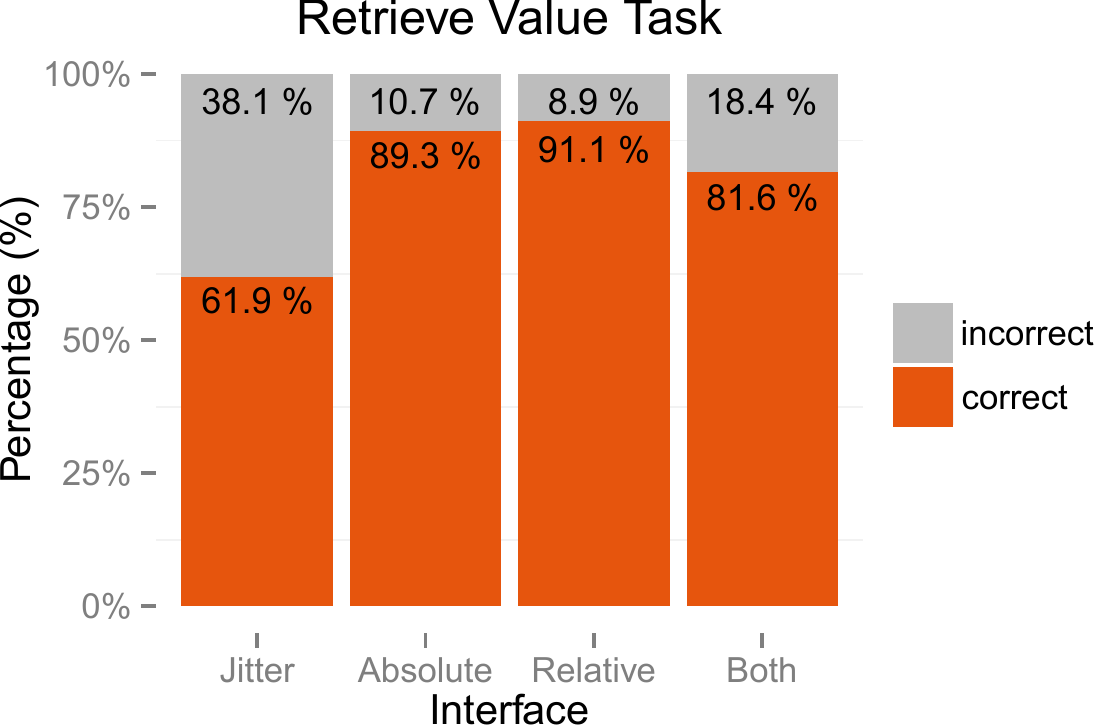}}
    \label{fig:none_task_type}
}
\subfigure[]{
    {\includegraphics[width=0.8\columnwidth]{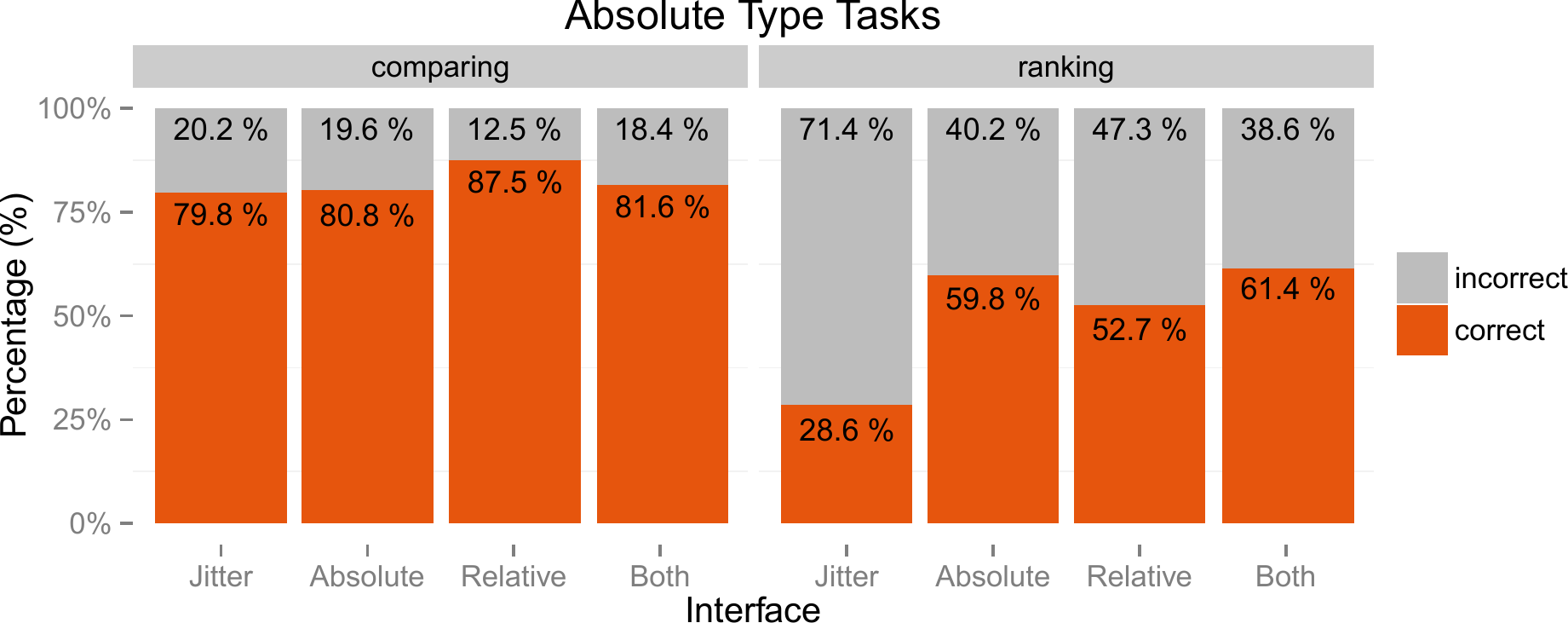}}
    \label{fig:absolute_task_percentage}
}
\subfigure[]{
    {\includegraphics[width=0.8\columnwidth]{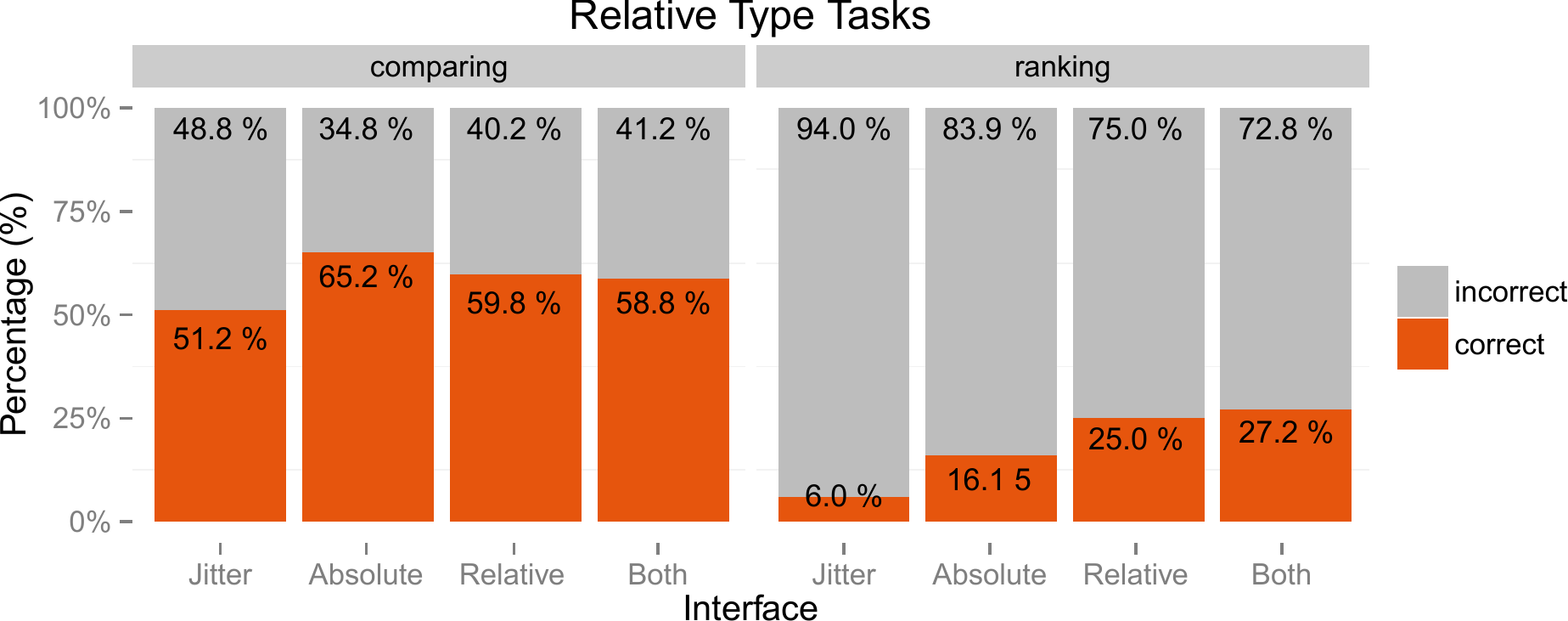}}
    \label{fig:relative_task_percentage}
}
\caption{(a) The percentage of participants who have got the answer correct for retrieving value task. (b) The percentage of participants who have got the answer correct for absolute type tasks for comparing and ranking. (c) The percentage of participants who have got the answer correct for relative type tasks for comparing and ranking.}
  \label{fig:correct_type_task}
\end{figure}

The number and percentage of participants who answered correct and incorrect answers are shown in Figure~\ref{fig:correct_type_task}.
Eventually, we had 42 participants for jitter, 56 participants for absolute, 56 participants for relative, and 57 participants fro both mode.

As the measure for each question was either correct or incorrect, a logistic regression was employed using PROC LOGISTICS in SAS.
For the retrieving-value task (Type 1), both the the absolute view and relative view had significant main effects (Wald Chi-Square = $18.58$, $p < 0.01$, Wald Chi-Square = 21.05, $p < 0.01$, respectively) with a significant interaction effect (Wald Chi-Square = 19.53, $p = 0.03$) (H1 confirmed).
For absolute-value tasks (Type 2 and 3), both the the absolute view and relative view had significant main effects (Wald Chi-Square = 10.35, $p < 0.01$, Wald Chi-Square = 10.35, $p  < 0.01$, respectively) with a significant interaction effect (Wald Chi-Square = 4.31, $p = 0.03$) (H2 confirmed).
For relative-value tasks (Type 4 and 5), only the relative view had a significant effect (Wald Chi-Square= 5.10, $ p = 0.02$ ) (H3 confirmed).

\subsubsection{Time spent}
The time spent (in seconds) for each question was compared using mixed-model ANOVA with repeated measures.
For the retrieving-value task, on average, the time spent (sec) for each interface was for jitter (44.26), absolute (56.84), relative (52.45), and both (56.57).
There was no significant difference between interfaces ($p > 0.05$ for all cases).

For  the absolute-value task (Type 2 and 3), on average, the time spent (sec) for each interface was for jitter (30.74), absolute (32.3), relative (33.6), and both (47.91).
The interface had a significant main effect ($F(3, 207) = 11.5, p <0.01$).
However, when we conducted pairwise comparisons with adjusted p values using simulation, the only significant difference in time spent was when using the both interface which took longer ($p< 0.01$ for all comparisons).

For relative-value task (Type 4 and 5), on average, the time spent for each interface was for jitter (26.6), absolute (31.12), relative (31.38), and both (46.78).
The interface had a significant main effect ($F(3, 207) = 10.12, p <0.01$).
However, when we conducted pairwise comparisons with adjusted p values using simulation, the only significant difference in time spent was when using the both interface which took longer ($p< 0.01$ for all comparisons).

\subsubsection{Confidence}

The 7-point Likert-scale rating was used for the level of confidence on their estimation.
For the value-retrieving task (Type 1), Kruskal-Wallis non-parametric test revealed that the type of interface had significant impact on the confidence level ($\chi^2(3) = 74.57 p < 0.01$).
The mean rating for each interface was for jitter (4.8), absolute (6.3), relative (6.0), and both (6.25).
A post-hoc Pairwise Wilcoxon Rank Sum test was employed with Bonferroni correction to adjust errors.
The jitter interface was significantly lower than the other three modes ($p<0.01$ for all cases).
There was no difference between absolute, relative, and both interfaces.

For absolute-value tasks (Type 2 and 3), Kruskal-Wallis non-parametric test revealed that the type of interface had significant impact on the confidence level ($\chi^2(3) = 18.32, p < 0.01$).
The mean rating for each interface was jitter (5.4), absolute (5.7), relative (5.0), and both (5.8).
A post-hoc Pairwise Wilcoxon Rank Sum test was employed with Bonferroni correction to adjust errors.
The interface with both mode was significantly higher than relative and jitter mode ($p<0.01$ for both), however no difference with the absolute mode.
The interface with absolute mode was significantly higher than relative and jitter mode ($ p<0.01$).

For relative-value tasks (Type 4 and 5), Kruskal-Wallis non-parametric test revealed that the type of interface did not have significant impact on the relative tasks ($\chi^2(3) = 4.1, p = 0.2$).
The mean rating was jitter (4.7), absolute (4.9), relative (4.9), and both (4.8).

One possibilities for result is that relative task might be harder than others.
The low correct percentage of questions are also shown in Figure~\ref{fig:relative_task_percentage}.
To see that, we have tested the confidence level among task types.
Kruskal-Wallis non-parametric test revealed that the type of task had significant impact on the confidence level ($\chi^2(2) = 148.1, p < 0.01$).
The mean rating for retrieving value (5.9), absolute (5.5), and relative (4.8).
The post-hoc Pairwise Wilcoxon Rank Sum test was employed with Bonferroni correction to adjust errors showed that all three task types have significantly different ($p <0.01$ for all cases).

\section {Discussions with Experts Feedback}

While the performance of executing low level unit task can explain functional part of new visualization, there are more qualitative aspects, such as aesthetics or playfulness.  Also the complex interaction techniques and features make it practically difficult to design test with statistical validation.  For this reason, complimentary evaluation of visualization technique with experts can be used~\cite{elmqvist2013}. To facilitate structured evaluations, a set of tasks and questions were given.  The experts were asked to follow the task and write their feedbacks in the discussion board.  They could see other people's comments. After each sessions, we organized feedbacks by moving to an existing thread or creating new topics.

In addition to this, authors requested the on-line data visualization and visual analytics community for opinions.  The intention is getting an initial response for adoption.  For they are voluntary and free from conflicts of interests, their response can be valuable for general adoption of a new visualization technique.

Because the feedbacks deal with advanced features and design choices, this session is combined with the discussion of issues triggered by the expert reviews.

\subsection {Expert Reviews}

We were able to get opinions from two graduate students and one professor whose major field is an information visualization.  Two sessions with graduate students were conducted in lab environments, where one of authors was available to answer quick questions or provide feedback if necessary.  But in general they were asked to follow the on-line guidelines and use discussion board to leave feedbacks.  They took about 70 minutes to finish the reviews. The professor was on his own while reviewing.  Their original responses were archived and available in the demonstration website at \url{http://www.gatherplot.org}.

The responses were positive in general, especially about the aesthetics and the layouts.  However many in-depth issues were discovered.  Most frequently pointed problems was the difficulty associated with the task of comparing absolute numbers of subgroups between groups of different size.  Especially comparison between a large percent subgroup in small group and a small percent subgroup in a large group is difficult.  For example, estimating  whether the second class female or third class male passengers survived more or not using figure~\ref{fig:aspectRatio} (a) and (b) is difficult.  The fundamental reason why this is difficult in gatherplots are because the areas are less effective than the length for perception~\cite{cleveland1984graphical}.  This task is well supported by the layout shown in figure~\ref{fig:lengthBasedLayout}~\ref{fig:genderVSAge}, which was dropped during the design process.   However the experts also provided other plausible suggestions to handle better, such as tool-tips, which shows the number of counts in the group.  One interesting solution was using the size of small groups as a mask for the anchor box, which will overlaid over the larger groups, so that the size estimation becomes easier.

 One interesting suggestion was changing to bar chart or pie chart.  For example, when there are only a few items in groups, due to large size of items, the estimation at relative mode can become inaccurate. During the design process, authors implemented this transform, where the rectangle shapes becomes thinner to become lines.  However the support for this mode was dropped later, because this mode loses sense of individual entities.

One reviewers suggested a subtle transition where the bin size changes incrementally by small steps to help maintain the sense of object constancy.

Relative mode was commented to be useful to understand the Bayesian inference problem, while one reviewer mentioning the difficulty of getting right setup with only small number of options.  Once correct setting is applied, it helped understanding of a counter intuitive result.  Also stretched out rectangles was helpful for user reminding that relative view is applied.

\subsection {Community Feedback}
During 3 days, it received comments from the 7 experts of data visualization and visual analytics community. All of them were positive.  Interesting remarks are following: one commented that he has been looking for a tool to create this for a long time.  other requested an open source library for gatherplots and ability to test their own datasets.  These comments imply that there may be a demand for technique in general.   Finally one pointed that this to be a general tool, which can be a standard requirements.

\section{Conclusion and Future Work}

We have proposed the concept of the gather transformation, which
enables space-filling layout without overdrawing while maintaining
object constancy.
We then applied this transformation to scatterplots, resulting in
gatherplots, a generalization of scatterplots, which enable overview
without clutter.
While gatherplots are optimal for categorical variables,
it can also be used to ameliorate overplotting caused by continuous
ordinal variables.
We discussed several aspects of gatherplots including layout,
coloring, tick format, and matrix formations.
We also evaluated the technique with a crowdsourced user study showing
that gatherplots are more effective than the jittering, and absolute
and relative mode serves specific types of tasks better.
Finally, in-depth feedback from an expert review involving
visualization reviewers revealed several limitations for the
gatherplots technique.
We addressed these weaknesses and suggested possible remedies.

We believe that gathering is a general framework to
formulate the transition of overlapping visualization to space-filling
visualization without sense of individual objects.
In the future, we plan on studying the application of this framework
to other visual representations to explore novel visualizations.
For example, parallel sets can be reconstructed to render individual
lines instead of block lines, which would enable combining both
categorical and continuous variables.
Gathering also enables mixing nominal variables and ordinal
variables in a single axis.
This can be pursued further, for example in a gathering lens that
gathers underlying objects according to a data property.
If we apply this lens to selected boundary in crowded region of
scatterplots, the underlying distribution of that region can be
revealed.

\bibliographystyle{abbrv}
\bibliography{gatherplot}

\end{document}